\documentclass{aa}
\usepackage{natbib}
\usepackage{graphicx}

\newcommand{\porb}{6$^{\rm d}$.724333}

\begin{document}

   \title{Multiperiodicity, modulations and flip-flops in variable star light curves}
   \subtitle{II. Analysis of II Peg photometry during 1979--2010}
   \authorrunning{M. Lindborg et al.}
   \titlerunning{II Peg photometry}

   \author{M. Lindborg
          \inst{1}
          \and
          M. J. Mantere
          \inst{1,2}
          \and
          N. Olspert
          \inst{3}
          \and
          J. Pelt
          \inst{3}
          \and
          T. Hackman
          \inst{1,4}
          \and
          G. W. Henry
          \inst{5}
          \and
          L. Jetsu
          \inst{1}
          \and
          K. G. Strassmeier
          \inst{6}
          }

   \offprints{Marjaana Lindborg\\
          \email{marjaana.lindborg@helsinki.fi}
          }   

   \institute{
     Department of Physics, PO Box 64, FI-00014
              University of Helsinki, Finland 
     \and
     Aalto University, Department of Information and Computer Science, 
              PO Box 15400, FI-00076 Aalto, Finland 
     \and
     Tartu Observatory, 61602 T\~{o}ravere, Estonia
     \and
     Finnish Centre for Astronomy with ESO (FINCA), University of Turku, 
     V\"{a}is\"{a}l\"{a}ntie 20, FI-21500 Piikki\"{o}, Finland 
     \and
     Center of Excellence in Information Systems, Tennessee State University, 
     3500 John A. Merritt Blvd., Box 9501, Nashville, TN 37209, USA
     \and
     Leibniz-Institute for Astrophysics Potsdam, An der Sternwarte 16, 
     14482, Potsdam, Germany
       }

\date{Received / Accepted}

\abstract{}
{
  According to earlier Doppler images of the magnetically active
  primary giant component of the RS CVn binary II Peg, the surface
  of the star was dominated by one single active longitude that was
  clearly drifting in the rotational frame of the binary system during
  1994-2002; later imaging for 2004-2010, however, showed decreased
  and chaotic spot activity, with no signs of the drift pattern. Here
  we set out to investigate from a more extensive photometric dataset
  whether such a drift is a persistent phenomenon, in which case it
  could be due to either an azimuthal dynamo wave or an indication of
  the binary system orbital synchronization still being incomplete.
}
{
We analyse the datasets using the Carrier Fit
  method (hereafter CF), especially suitable for analyzing time series
  in which a fast clocking frequency (such as the rotation of the
  star) is modulated with a slower process (such as the stellar
  activity cycle).
}  
{
  We combine all collected photometric data into one single data set,
  and analyze it with the CF method. As a result, we confirm the
  earlier results of the spot activity having been dominated by one
  primary spotted region almost through the entire data set, and the
  existence of a persistent, nearly linear drift. Disruptions of the
  linear trend and complicated phase behavior are also seen, but the
  period analysis reveals a rather stable periodicity with $P_{\rm
    spot}=$6$^{\rm d}$.71054$\pm$0$^{\rm d}$.00005. After the linear
  trend is removed from the data, we identify several abrupt phase
  jumps, three of which are analyzed in more detail with the CF method. These
  phase jumps closely resemble what is called flip-flop event,
  but the new spot configurations do not, in most cases, persist for
  longer than a few months.
}
{
  There is some evidence of the regular drift without phase jumps
  being related to the high state, while complex phase behavior and
  disrupted drift pattern to the low state of magnetic activity. These
  findings do not support the scenario of the objects' rotational
  velocity still being de-synchronized from the orbital period of the
  binary system. Neither is any straightforward explanation provided by
  the azimuthal dynamo wave scenario, as dynamo models preferentially
  show solutions with stable drift periods.  }
\maketitle

\section{Introduction}

II~Peg is one of the most extensively studied SB1 RS CVn systems;
recently, surface temperature
\citep[e.g.][]{Sveta98b,Sveta99,Gu03,lindborg11,Hackman12} and
magnetic field maps have been published \citep[][]{Oleg12,Carrol2007,Carrol2009}. The star's
photometric light curves have also been intensively studied with time
series analysis and spot modelling methods \citep[see
e.g.][]{Henry95,Rodono00,Hackman11,Roettenbacher11}, 
all the light
curve variations being commonly interpreted being due to starspots
\citep[see e.g. the recent review by][]{strassmeier2009}.
One peculiar
feature apparent especially from the paper by \citet{lindborg11} is
the tendency of the magnetic activity to cluster on one primary active
longitude, the position of which is linearly drifting in the orbital
frame of reference of the binary system. 
Clustering of the spot activity on one single longitude has also been
reported in other short-period eclipsing RS CVn systems e.g. by
\citet{Zeilik}.
In some earlier studies of II Peg \citep[e.g.][]{Sveta99} indications
of flip flops, i.e. the activity abruptly jumping from one longitude
to another, separated roughly by 180 degrees, were reported, but the
results of \citet{lindborg11, Hackman12} did not give support to this
interpretation. 

Comparing with
theoretical dynamo models \citep[see
e.g.][]{KR80,Radler75,Moss95,MKK99,Ilkka02}, the drift was
preliminarily associated with an azimuthal dynamo wave of the
non-axisymmetric dynamo solution, resulting from dynamo action due to
rotationally influenced turbulent convection without non-uniform
rotation (so called $\alpha^2$ dynamo mechanism). Typically, the
dynamo models predict dynamo waves either rotating faster or slower
with a {\it constant} angular frequency over
time. In addition to the dynamo wave changing its position in the
phase-time plot, the solutions may exhibit oscillatory behavior
concerning the energy level of the mean magnetic field, reminiscent of
a stellar cycle. In a recent study of \citet{Mantere13} the modelling
showed that the drift cycle length (i.e. the time required for the
non-axisymmetric magnetic structure to make a full rotation in the
rotational frame of reference) and the oscillation period of the
magnetic energy were always connected to each other. It has to be
noted that in this particular study, only a very small fraction of the
parameter space was mapped, within which solutions for which the drift
cycle length and oscillation of the magnetic energy levels coincided
were found to be preferred.
Flip-flop type switches have been found in dynamo models, where
relatively small amounts of differential rotation have been allowed
for \citep[so called $\alpha^2 \Omega$ modes, see
e.g.][]{Elstner2005,Korhonen2011}. In these models, flip-flops result
from the competition of oscillatory axisymmetric mode and the stable
non-axisymmetric mode of comparable strengths.

II Peg belongs to a binary system, the rotational speed of which can
be deduced from radial velocity measurements
\citep[e.g.][]{Sveta98a}. It is well known that the tidal effects due
to the presence of the companion star will circularize the orbit of
the binary system and force the rotation of the star to be
synchronised to the orbital period of the binary in a time frame of a
billion years after settling to the main sequence \citep[see
e.g.][]{Zahn89}. In the case of the rotation of the
star not yet being fully synchronised, i.e. the rotational speed still
being larger than implied by the orbital period of the binary, one
would expect a systematic drift of the lightcurve minima when wrapped
with the binary period; this constitutes the second possible
explanation to the drift of the spots on the star's surface. If this
was the case, a persistent drift of the activity tracers, making the
stellar surface and rotation visible to the observer, over the whole
extent of the time series should be seen - the amplitude of the
spottedness or the magnetic field strength could still vary, but the
drift should still be visible if this amplitude variation was
removed. In contrast to the dynamo wave scenario presented above, no
obvious connection between the drift versus the modulation should be
visible.

Intriguingly, the drift was no longer visible in the Doppler
(hereafter DI) and Zeeman Doppler images (hereafter ZDI) of
\citet{Hackman12} and \citet{Oleg12}, who also reported on the clear
diminishing trend in the spottedness and magnetic field strength of
the object. This was interpreted as the star's activity declining
towards its magnetic minimum, due to which the spot distribution on
the stellar surface was postulated to be more stochastic than during
the high activity state, veiling the possible drift due to the
mechanism producing it underneath.  We note that the results of
\citet{Carrol2007,Carrol2009}, computed from the same data with a
different ZDI inversion method \citep{carroll2009b}, also indicate a
reduction in the magnetic field strength between the observing seasons
2004 and 2007. They, however, report that a dominant and largely
unipolar field seen in 2004 changed into two distinct, large-scale,
bipolar structures in 2007, not completely consistent with the
proposed picture of the star declining towards its magnetic minimum
with less pronounced, stochastic spot activity.

Based on these analyses, however,
it is not possible to fully rule out the scenario of the drift having
really disappeared. A time-dependent drift period would not match either
of the two proposed scenarios, and would be an important finding as
such.

  Whichever the cause or persistence of the drift, this type of a
  problem is especially suited to be analyzed with the recently
  developed Carrier Fit (CF) analysis method \citep{Pelt11}, which
  is based on the idea of continuous fitting of the time series
  (e.g. a light curve of a star) clocked by a certain shorter carrier
  period (or larger frequency) while being modulated by a longer
  period (or smaller frequency). This method has already successfully
  been applied to analyze the phase changes seen in the light curve of
  FK Com \citep{Hackman13}. In the case of II Peg, the orbital
  period of the binary system serves as the first guess of the carrier
  period, while the apparent modulation of the amplitude of the spots
  and/or magnetic field present the modulation period. If
  either of the proposed scenarios (azimuthal dynamo wave, orbital
  de-synchronization) would be true, we should be able to find a
  persistent drift period. The two mechanisms could be differentiated
  from each other by examining the correlation of the modulating
  period to the carrier; while the dynamo scenario might show up as
  these two being connected, the other scenario would imply no
  connection. Another method with which these hypothesis could be tested
  would be to use a 'local' period search method with a sliding
  window, such as the CPS method of \citet{Lehtinen11}, which would
  allow the comparison of the mean active longitude rotation period to
  the short-term photometric period.

  In this paper, we aim to investigate the drift phenomenon in detail,
  combining all the available photometry of the star into our
  analysis. The primary goal of this paper is to study whether a
  systematic drift period persisting over time exists with the CF
  method. In Sect.~\ref{datasect} we introduce the data used, in
  Sect.~\ref{CF} present a short summary of the CF method, in
  Sect.~\ref{res} present our results, in Sect.~\ref{comp} compare the
  photometric results to earlier Doppler images, and in
  Sect.~\ref{concl} summarize and discuss the possible implications of
  the results obtained.

\section{Photometric datasets}\label{datasect}

\begin{figure*}
\begin{center}
\includegraphics[width=\textwidth]{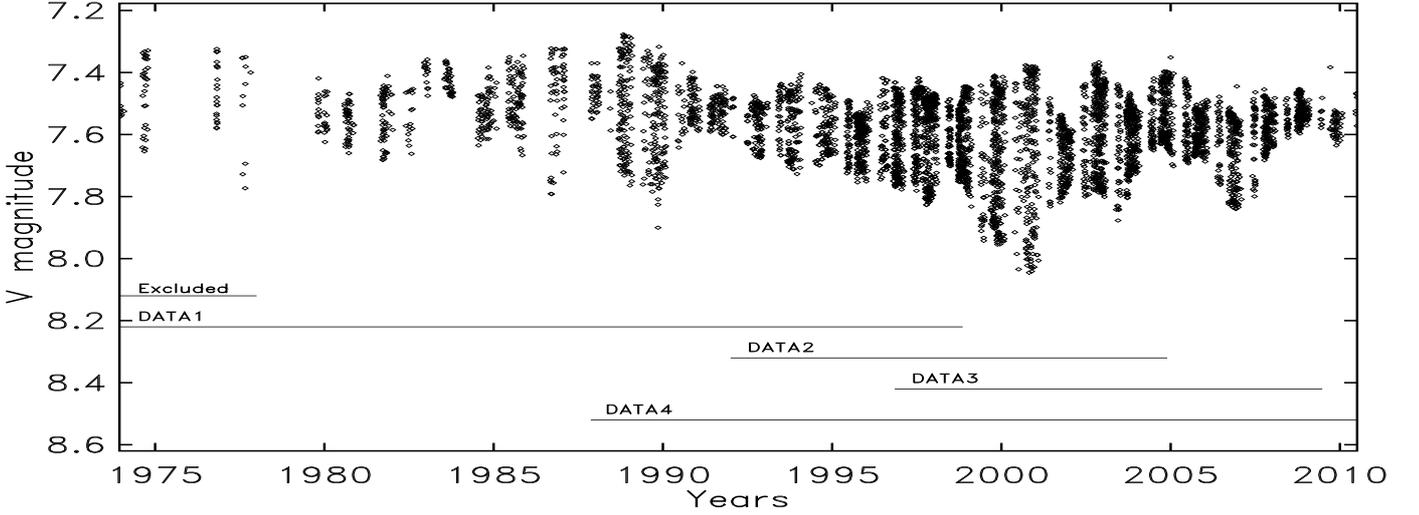}
\caption{All the photometric data normalized and combined.}\label{data}
\end{center}
\end{figure*}

We use four different data sets for our CF analysis. The first data
set (hereafter DATA1), published and analyzed by \citet{Rodono00},
covers the years 1973-1998. The second data set (hereafter DATA2) was
published by \citet{Messina08} covering the years 1992-2004, therefore
partly overlapping with DATA1. The third data set (hereafter DATA3)
consists of unpublished observations with the Wolfgang-Amadeus, the
university of Potsdam/Vienna twin automatic photoelectric telescope
(APT), covering the years 1996-2009, again partially overlapping with
the previous datasets. The fourth dataset (hereafter DATA4) was
obtained with the Tennessee State University T3 0.4-m Automated
Photometric Telescope at Fairborn Observatory in Arizona, covering the
time span of 1987-2010 \citep[see also][for another analysis of the
same dataset]{Roettenbacher11}. These datasets are combined, to
collect as extensive a dataset as possible, with the densest possible
coverage of observations.  The combination was done by shifting the
magnitudes and rescaling the amplitudes to get the best match in
overlapping areas. The quality of the match was measured by computing
mean-squared differences between magnitudes in data point pairs where
the time distance was less than 0.2 days. The optimal scaling shift
and magnitude corrections were obtained by a least-squares
minimization. The first six years of DATA1 (1973-1979) have gaps in
time too wide for the CF analysis so these data were removed from the
combined data set (see Fig.~\ref{data}). We also divided the combined
dataset into shorter segments, either to look for phase jumps or
flip-flop type events, following the ideas presented in
\citet{Hackman13}, or to be able to identify the best period
describing the drift of the primary light curve minimum in the orbital
frame of reference.

\section{The Carrier Fit method}\label{CF}

Astronomical data is complicated for a straightforward Fourier
analysis, because the observed light curves and spectra usually have
long time gaps between observations. This makes the analysis difficult
and creates false peaks due to the regularities in the gap structure
of the input data \citep[see e.g. the review by][and references
therein]{SC03}. \citet{Pelt11} developed a novel method for stellar
light curve analysis, based on a simple idea of decomposing the light
curve into two separate components: a rapidly changing carrier tracing
the regular part of the signal, and its slowly changing modulation.
The carrier frequency can be obtained from observations (e.g. the
period can be obtained from rotational velocity) or estimated from the
data as a mean frequency. The smooth modulation curves are
described by trigonometric polynomials or splines. For a complete
description of the method we refer to \citet{Pelt11} and as another
practical application for stellar light curves to \citet[][analysis of
FK Com]{Hackman13}; here we merely highlight the most important input
quantities and properties of the method.

The proposed composition of the light curve is described with the
following formula:
\begin{equation}
f(t) = a_{0}(t) + \Sigma^{K}_{k=1} (a_{k} (t) \cos (2 \pi t k\nu_{0} ) + b_{k}(t) \sin (2 \pi t k \nu_{0} )),
\end{equation}
where $\nu_0$ is the carrier frequency, $a_0(t)$ is the time-dependent
mean level of the signal, $K$ is the total number of harmonics
included in the model, describing the overtones of the basic carrier
frequency, while $a_k(t)$ and $b_k(t)$ are the low-frequency signal
components. In the case of II Peg, the first guess of the carrier
frequency $\nu_0=P_0^{-1}$ comes from the consideration of the system
being a binary system, most likely with synchronized orbits of the
components. Therefore, we adopt the orbital period of the binary
system, $P_0$=\porb, estimated by \citet{Sveta98a} as the
first carrier. The next task of the analysis is to choose a suitable
model for the modulators. In \citet{Pelt11} we proposed two different
types of models based on either trigonometric or spline
approximation. The II Peg light curve contains fairly small amount of
complicated features, making the usage of the trigonometric model
adequate.

We build the trigonometric modulator model in the following way: Let
the time interval $[t_{\min},t_{\max}]$ be the full span of our input
data, based on which we define a data period $D = C\times
(t_{\max}-t_{\min})$, where $C$ is a so called coverage factor, the
value of which must be larger than unity. Next we construct a
truncated trigonometric series using the corresponding data frequency,
$\nu_{D}= 1/D$, i.e.:
\begin{equation}
a(t) = c_0^a  + \sum\limits_{l = 1}^L {\big ( c_l^a\cos (2\pi tl\nu_D ) + s_l^a\sin (2\pi tl\nu_D )\big ),}
\end{equation}
and
\begin{equation}
b(t) = c_0^b  + \sum\limits_{l = 1}^L {\big ( c_l^b\cos (2\pi tl\nu_D ) + s_l^b\sin (2\pi tl\nu_D )\big ),} 
\end{equation}
where $L$ is the total number of harmonics used in the modulator
model. The process described by the modulators and data period $D$
must be slow, i.e. $D$ must be significantly longer than the carrier
period $P_0$. The data of each set span over several thousands of
days, due to which we conclude that 10000 days is an appropriate value
for the data period; the corresponding coverage factors are roughly
$C$=1.2, well in the range considered to be satisfactory
$C=1.1\ldots1.5$ \citep[see][]{Pelt11}.

Next, proper estimates for the expansion coefficients are computed for
every term in the series for the fixed carrier frequency $\nu_0$ and
$D$; this is a standard linear estimation procedure and can be
implemented using any standard statistical package \citep[see][for
detailed description]{Pelt11}. If all coefficients ($a_k,b_k$)
consist of the same number of harmonics $L$ and we approximate
separate cycles by a $K$-harmonic model, then the overall count of
linear parameters to be fitted is $N=(2 \times L+1)\times(2 \times
K+1)$. The optimal number of harmonics, $K$, depends on the
complexity of the phase curves. The choice of the optimal number of
modulator harmonics, $L$, is constrained by the longest gaps in the
time series. As the light curve of II~Peg appears not to be very
complicated, as discussed at length in Sect.~\ref{res}, we find $K=3$
and $L=12$ adequate for the analysis of the combined data set (Stage
1).  For the short segments the number of data points tends to be
small and therefore we choose $K=2$ (instead of $K=3$) and $L=3$ to
reduce the number of model parameters (Stage 3).

Finally, we have found it convenient to visualize our results in the
following way. We begin with calculating a continuous curve estimate,
$\hat y(t)$, from the randomly spaced data set containing gaps, using
a least-squares fitting scheme based on the carrier frequency. As a
result, we can also model the gaps between the data, which allows us
to get a smooth picture of the long-term behaviour. The continuous
curve is then divided into strips, the length of each being the
carrier period. Each data strip is then normalized with its extrema,
so that after the normalization the data spans the range of
$[-1,1]$. We note that without this normalization procedure, epochs of
weaker spot activity, showing up as less variation in the light curve,
would be drowned in the signal from strong spot activity
epochs. Finally, after normalization, we collect the strips along the
time axis. To facilitate the interpretation of the plots, we do not
restrict them to the interval of $[0,1]$ over phase, but extend our
phase axis to somewhat larger and smaller values. At the bottom of the
plots we show the time distribution of the actual observations in the
form of a ``bar code''. The black stripes are used to denote periods
during which at least one observation is available, while the bright
stripes represent gaps. This method of visualisation allows us to
verify that the model correctly fits into the data rather than into
the gaps.

\section{The CF analysis procedure and results}\label{res}

During the first stage of our investigation, we perform the CF
analysis for the combined dataset (Fig.~\ref{data}) as a whole. As
previous photometric and spectroscopic investigations have already
suggested, we find one single photometric minimum, i.e. one large
spotted region, dominating the light curve almost for the entire span
of the data. This region often exhibits a linear downward trend in the
phase diagram that is wrapped with the orbital period, indicating that
it is rotating somewhat faster than the orbital period of the binary
system. During some epochs, this trend is disrupted for some years,
but it appears again. Therefore, we set out to determine this period,
which can be thought to describe either the rotation of the spotted
region on the surface of the object or the signature of an underlying
magnetic structure feeding the surface; this constitutes the second
stage of our analysis. In the third stage, we locate all the epochs
during which significant deviations from the drift trend can be
seen. We zoom into these epochs, and perform a more detailed CF
analysis on them, to be able to conclude on the nature of the more
complex phase behavior occasionally seen on the object.

\subsection{Stage 1: analysis of the combined dataset}

\begin{figure}
\begin{center}
\includegraphics[width=\columnwidth]{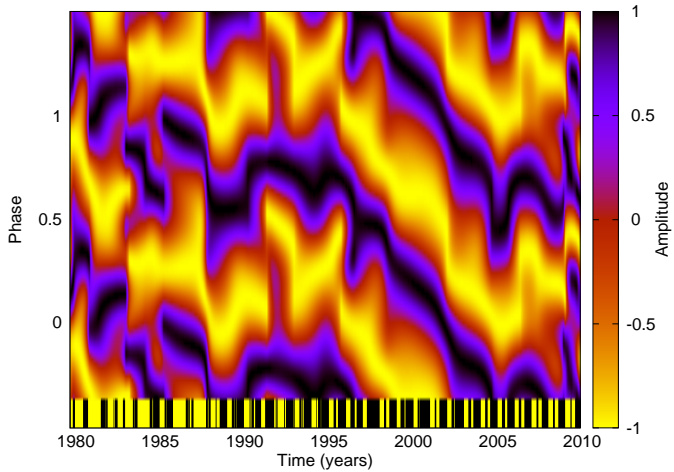}
\caption{Continuous fit of to the combined dataset from CF analysis
  with carrier $P_0$.}\label{ALL_DATA_CF}
\end{center}
\end{figure}

We begin by analyzing the whole combined dataset
[DATA1,DATA2,DATA3,DATA4], i.e. all the data points in Fig.~\ref{data}
excluding only the very early points with too large gaps during
1973-1978. The CF analysis results with $P_0$=\porb, $C$=1.15, number
of carrier harmonics $K$=3 and modulation harmonics $L$=12 is depicted
in Fig.~\ref{ALL_DATA_CF}.  The linear downward trend in the orbital
frame is well visible especially during the years 1995--2005
(subinterval of DATA3 and DATA4), and evidently present also during
the years 1979-1988 (subinterval of DATA1 and DATA2). In the early
part of the data, however, the phase of the principal light curve
minimum is alternating around the phase of the mean trend, with
seemingly regular timing. Moreover, the phase jumps seem to occur
rather abruptly.

During the time interval of 1988-1995, the linear trend is visibly
disrupted, manifested by the phase of the principal minimum staying
constant in the orbital frame for these years. In 1995, the linear
trend, however, resumes with a roughly equivalent period to the
earlier years. After 2005, a similar disruption is seen again, followed
by very complex behavior in the end of the data set.

In summary, four main features can be isolated from the first stage of
our analysis: 
\begin{itemize}
\item The data is dominated by one single light curve
minimum, i.e. one spotted region, almost the entire span of the data.
\item There is a linear downward trend with a period faster than \porb, 
visible during 1979-1988 and 1995-2005. 
\item Disruptions of the linear
trend occur during 1988-1995 and 2005-2010. 
\item Some abrupt phase jumps
are seen, not directly correlated with either the presence/absence of
the linear trend.
\end{itemize}

\subsection{Stage 2: finding the drift period}

The next task is to estimate the period that best
describes the linear trend seen in the dataset, $P_{\rm spot}$. As
already evident from the first stage of the analysis, the trend is not
visible at all times of the data, meaning that the period is not a
completely stable one.

Suitable methods to estimate light curves with significant deviations
from harmonic behavior are based on phase dispersion minimization. The
Stellingwerf method \citep{Stellingwerf} is among the best known
of such methods, and is used in the following as our base method for
period search. However, the method should be refined to be capable of
dealing with a periodicity that is not completely time independent and
stable. For this purpose a more general formulation of the phase
dispersion is needed. In this work we introduce a variant of the
statistic proposed earlier by \citet{Pelt1983} that is referred to as
{\it Method~2} in the remaining of the paper. Specifically, for each
trial period $P$ we compute the
dispersion of phases wrapped with this
particular period as
\begin{equation}
D(P) = \frac{{\sum\limits_{i = 1}^{N - 1} {\sum\limits_{j = i + 1}^N {g(t_i } } ,t_j ,P,\Delta t)[f(t_i ) - f(t_j )]^2 }}{{\sum\limits_{i = 1}^{N - 1} {\sum\limits_{j = i + 1}^N {g(t_i } } ,t_j ,P,\Delta t)}},
\end{equation}
where $g(t_i,t_j,P,\Delta t)$ is significantly greater than zero only when
\begin{eqnarray}
t_j  - t_j  &\approx& kP,k =  \pm 1, \pm 2, \ldots {\rm \ \ \ and}\\
\left| {t_j  - t_j } \right| &\le& \Delta t.
\end{eqnarray}
The first condition selects only the data point (magnitude) pairs
$[f(t_i)$,$f(t_j)]$ the phases of which are approximately the same and
the second condition restricts the time interval of the selected data
points to a certain pre-selected time range $\Delta t$. By computing
$D(P)$ for a sequence of trial periods we get the phase dispersion
spectra. The minima in these spectra indicate probable periodicities;
due to the normalization the expected $D(P)$ value for randomly
scattered phase diagrams is 2. For the particular case when the time
difference limit is longer than the full data span the obtained
spectra practically coincide with traditional Stellingwerf spectra. By
making the time limit $\Delta t$ shorter we can analyse phase scatter for
the case when different cycles match each other only locally; for the
present analysis, we find $\Delta t$ of 300 days optimal. In this case
the final minimum in the $D(P)$ spectrum indicates a certain estimate of
the mean cycle length (mean period). In this way we can analyse data
sets that are not coherent during the full data span but only
locally. 

It is reasonable to compute Stellingwerf and $D(P)=D(1/\nu)$ spectra
with a fixed step along the frequency $\nu$. The optimal step size,
\begin{equation}
\Delta \nu = { 0.05 \over t_{max}-t_{min} }, 
\end{equation}
limits the phase change from one trial to the next by an upper limit of $0.05$
for all pairs along the full time span $[t_{min},t_{max}]$. The
distance between neighbouring periods depends on the period argument $P$
\begin{equation}
\Delta P = P^2 \Delta \nu.
\end{equation} 
This value can roughly be used to characterize the precision of
the obtained period estimates.
 
The exact error estimates and significance levels for the estimated
mean periods can be computed by using either standard regression
techniques, Monte Carlo type methods, the Fischer randomization
technique or bootstrap. However, we note that these are only formal
error estimates. The real scatter of the mean period heavily depends
on the physics involved and to estimate it we need significantly
longer data sets than the ones available for this study. In our
particular case we used the standard regression analysis method to
estimate the errors of the final period. We iteratively refined the
obtained period and used a correlation matrix to compute the
dispersion of the refined period.

\begin{table}
\begin{center}
\begin{tabular}{ccccc}\hline
Method & Data & Period $P$[d] & $\Delta P$ \\ \hline
Stellingwerf &All &6.7108 & 0.0002 \\
Method~2      &All &6.7106 & 0.0002  \\
Stellingwerf &Subset &6.7109 & 0.0004 \\
Method~2      &Subset &6.7109 & 0.0004 \\ \hline
\end{tabular}\caption{Periods derived for combined and subsets of the data with two different methods.}\label{periods}
\end{center}
\end{table}

\begin{figure}
\begin{center}
\includegraphics[width=\columnwidth]{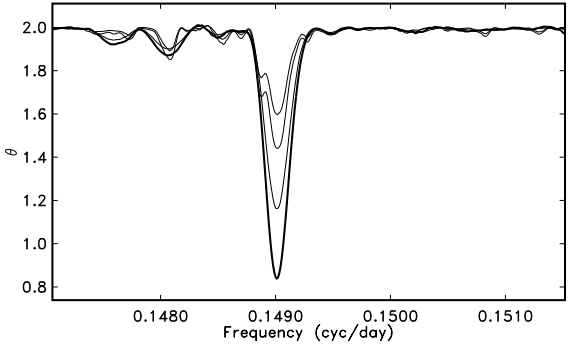}
\caption{Periodograms from two different methods with two different
  datasets.}\label{periodograms}
\end{center}
\end{figure}

The Stellingwerf periodogram for the full dataset, plotted with the
thinnest gray line in Fig.~\ref{periodograms}, shows the shallowest
minimum split into several peaks. The form of the spectrum does not
change when the correlation length is decreased using Method~2, but
the minimum gets deeper (2nd thinnest gray line). This is an
indication supporting the hypothesis that the period is not completely
stable. For the subset with the clearest linear trend seen in the
phase diagram, the spectrum is characterized with only one clear peak
in the periodogram with both methods (two thickest lines). Decreasing
the correlation length again using Method~2 makes the minimum deeper
(the thickest line).

\begin{figure}
\begin{center}
\includegraphics[width=\columnwidth]{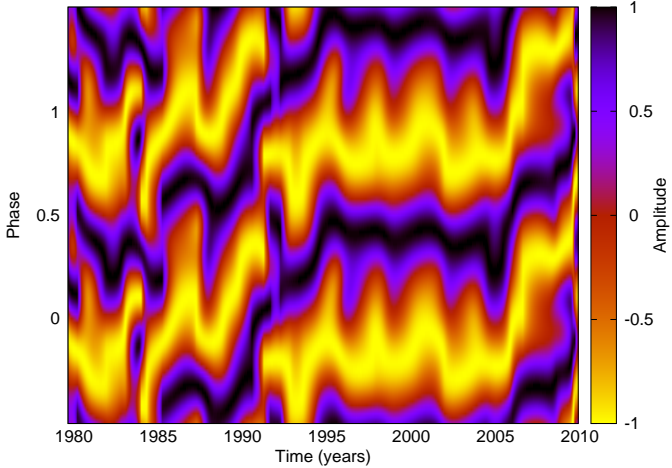}
\caption{Phase diagram with the new carrier, $P_{\rm spot}=$6$^{\rm d}$.71086.}\label{NEWCARRIER}
\end{center}
\end{figure}

Most importantly, the best period is practically the same for both
datasets analysed with either method. This is an indication of the
trend dominating in the data, despite of the disruptions and phase
jumps additionally seen in it occasionally. Therefore, we can pin down
the rotation period of the spotted region to be $P_{\rm
  spots}=$6$^{\rm d}$.71086$\pm$0$^{\rm d}$.00007, the value of the
period obtained after the least-squares refinement of the initial
estimate.  The phase diagram, wrapped with the new period, is shown in
Fig.~\ref{NEWCARRIER}. This figure reveals quite a different picture
from the phase diagram wrapped with the orbital period - the trend
having been removed, the primary minimum mainly stays on one straight
line at phase 0.5, although smooth variability up and down can be
seen. The phase changes around the mean drift period during the early
part of the data (1979-1990) now look fairly similar to the
fluctuations seen in the later part with a clear trend
(1995-2005). Abrupt phase jumps are more easy to identify; three major
events can be seen during 1984-1985, 1990-1991, and 2009-2010. Next,
we concentrate on investigating the phase jumps in more detail.

\subsection{Stage 3: analysis of short segments}

\begin{figure*}
\begin{center}
\includegraphics[width=0.33\textwidth]{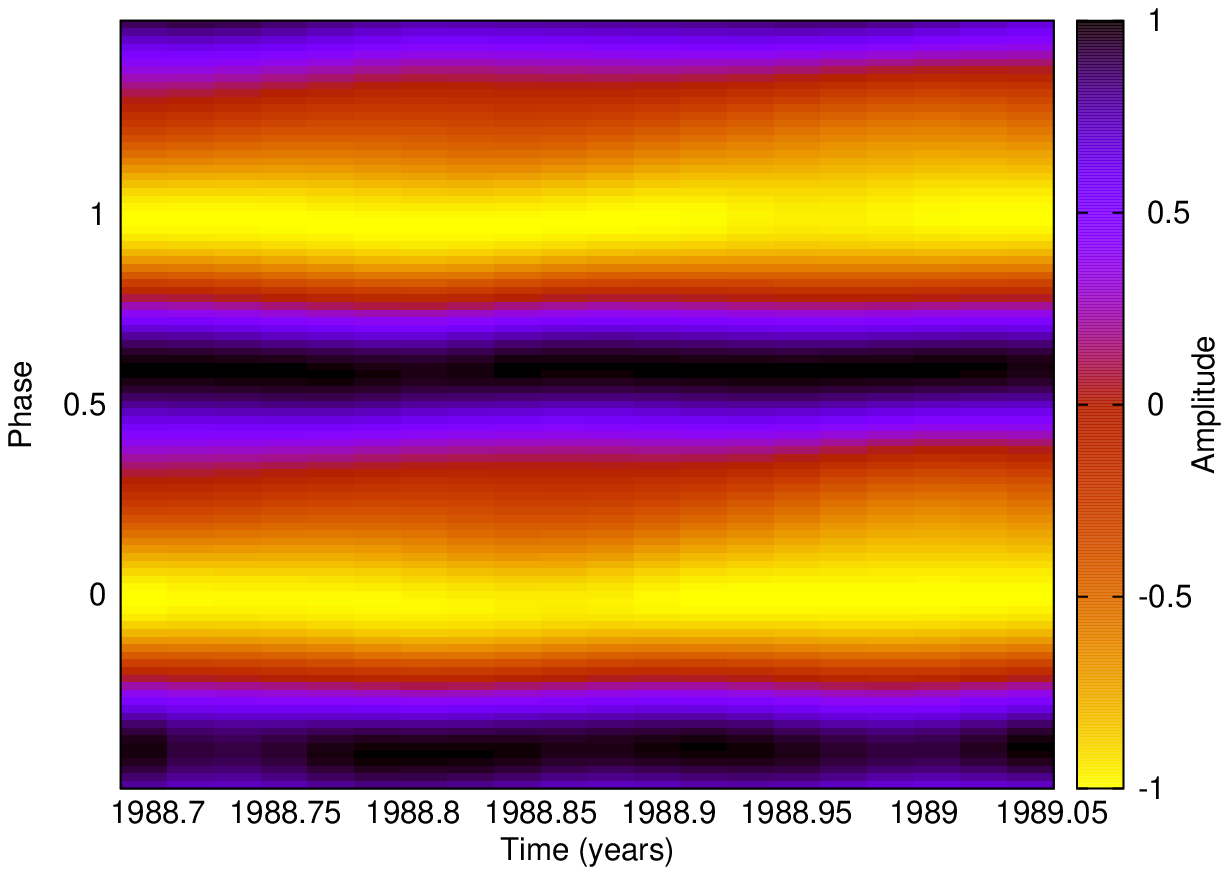}\\
\includegraphics[width=0.33\textwidth]{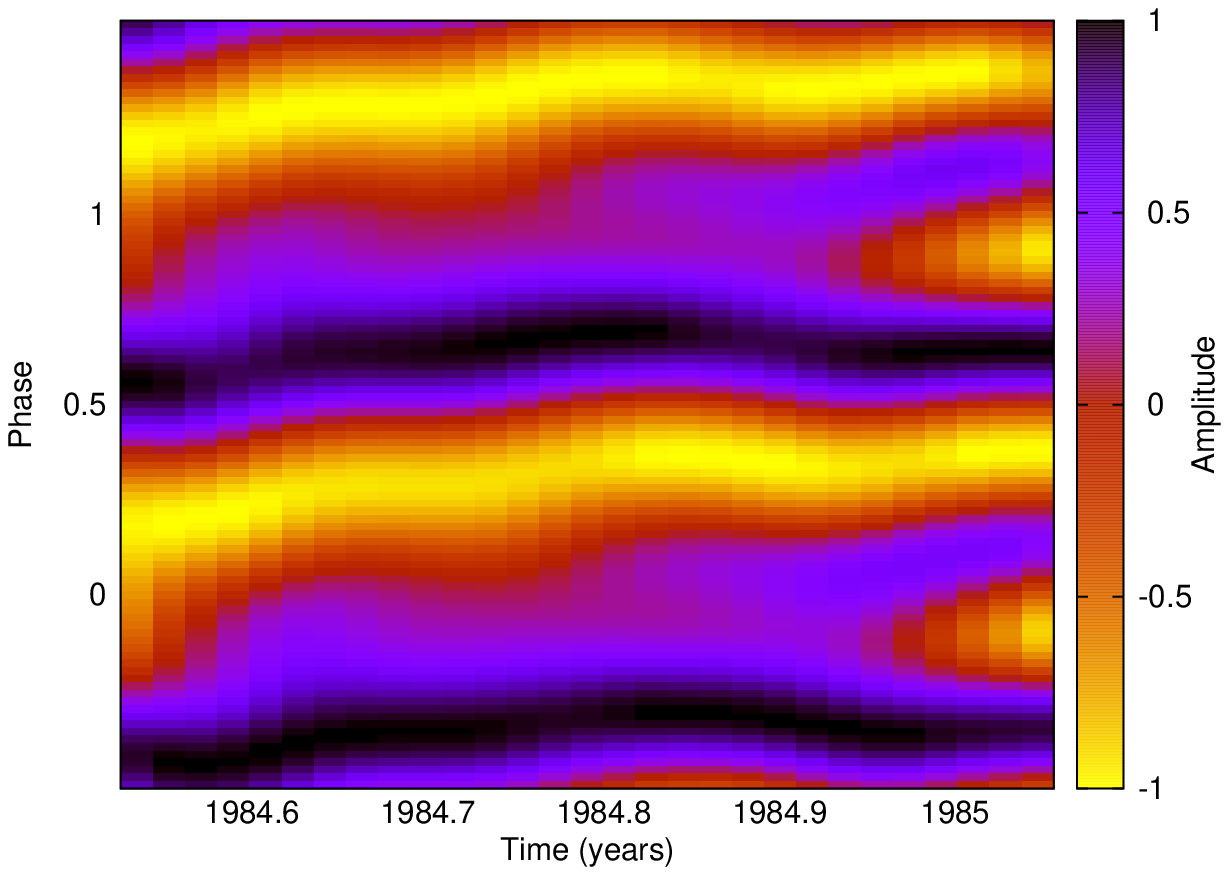}\includegraphics[width=0.33\textwidth]{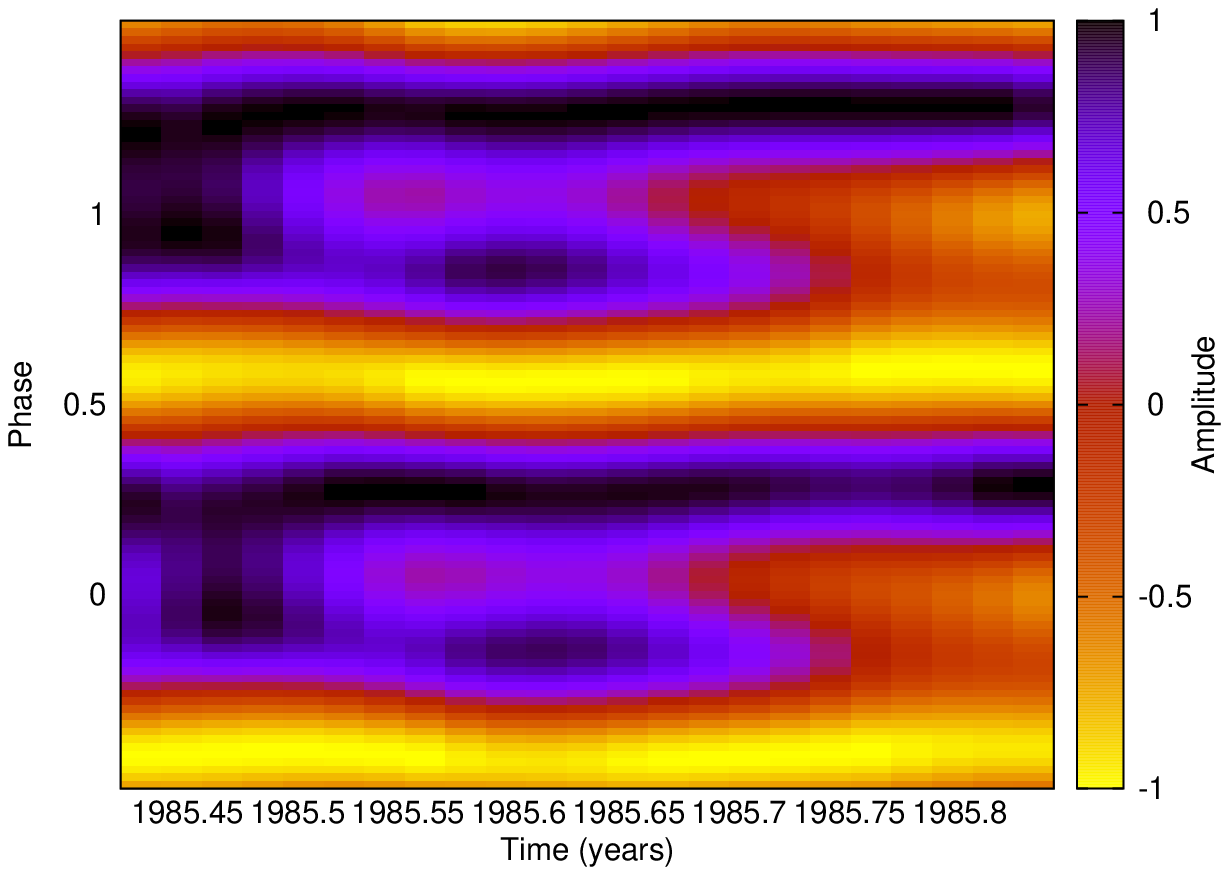}\includegraphics[width=0.33\textwidth]{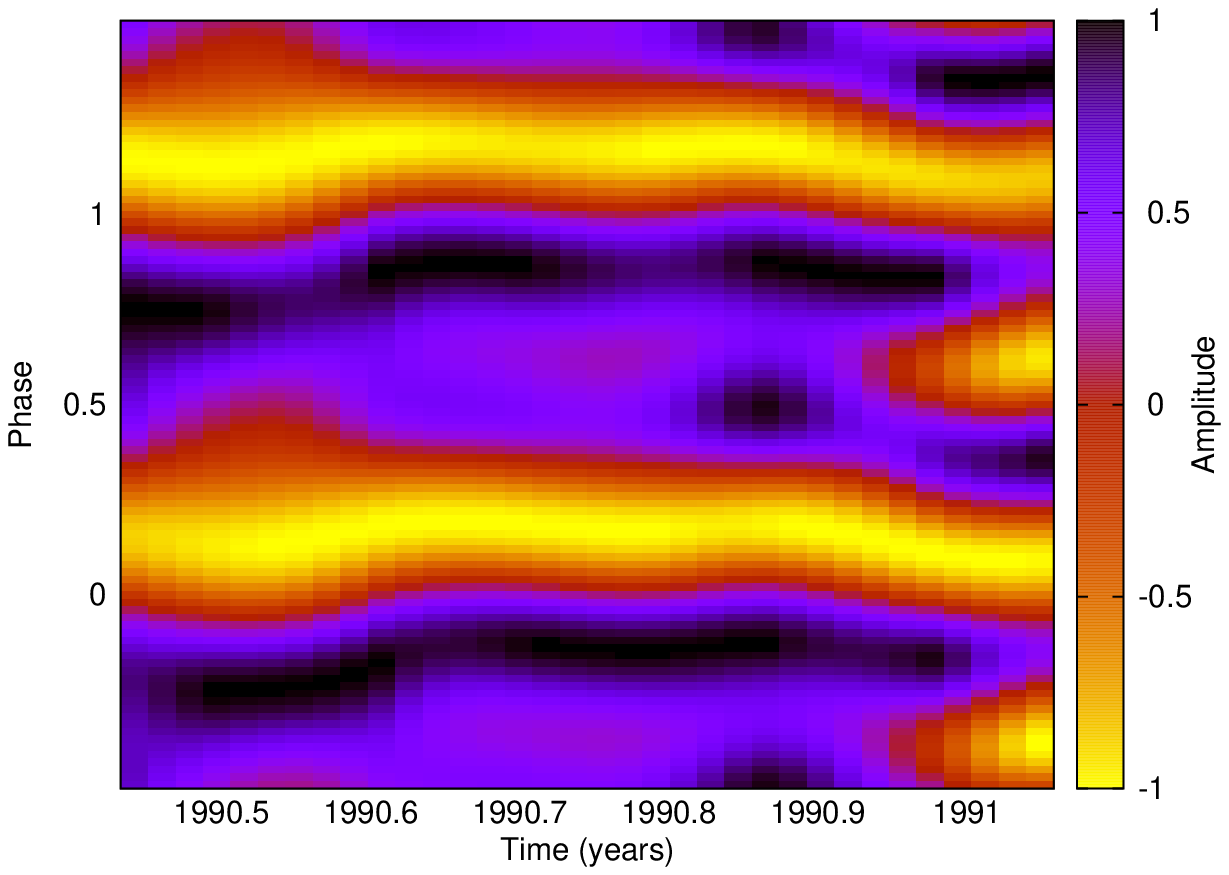}\\
\includegraphics[width=0.33\textwidth]{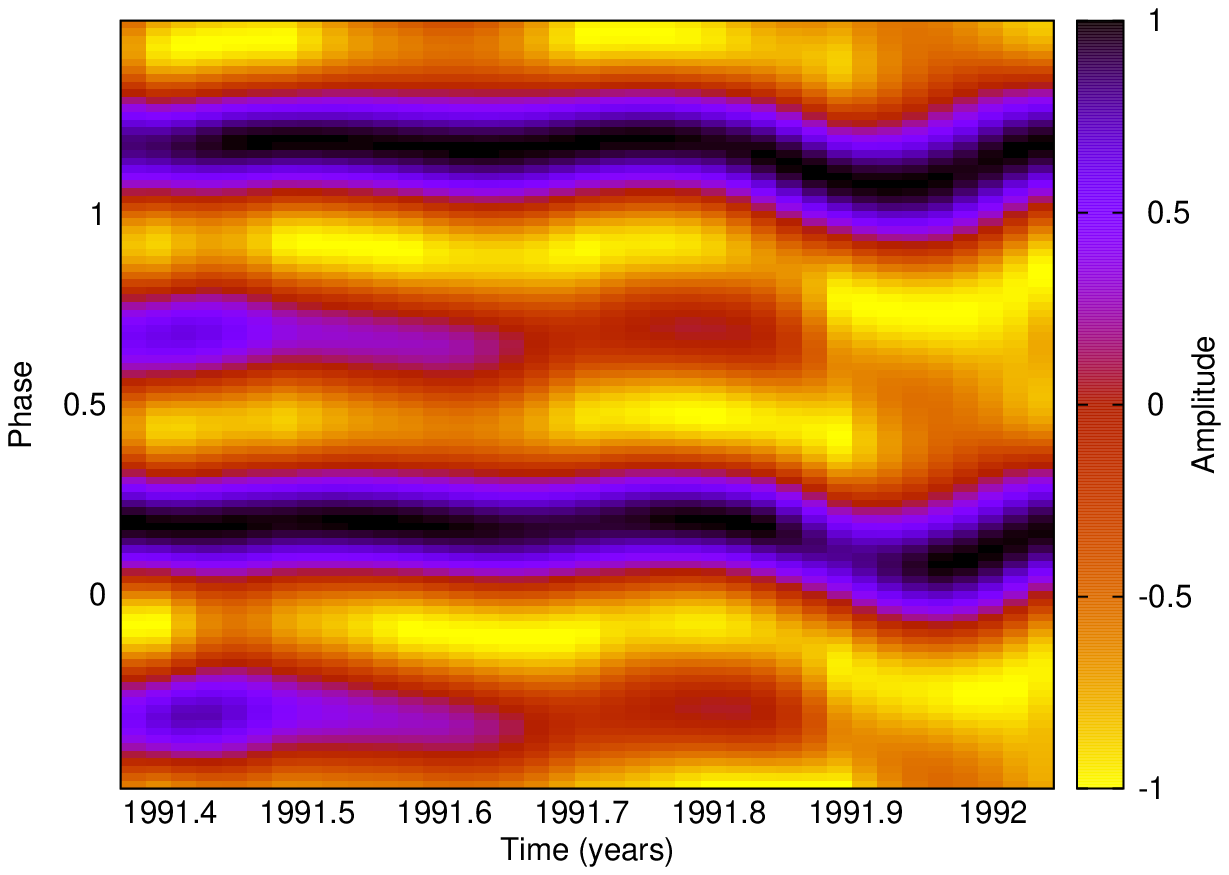}\includegraphics[width=0.33\textwidth]{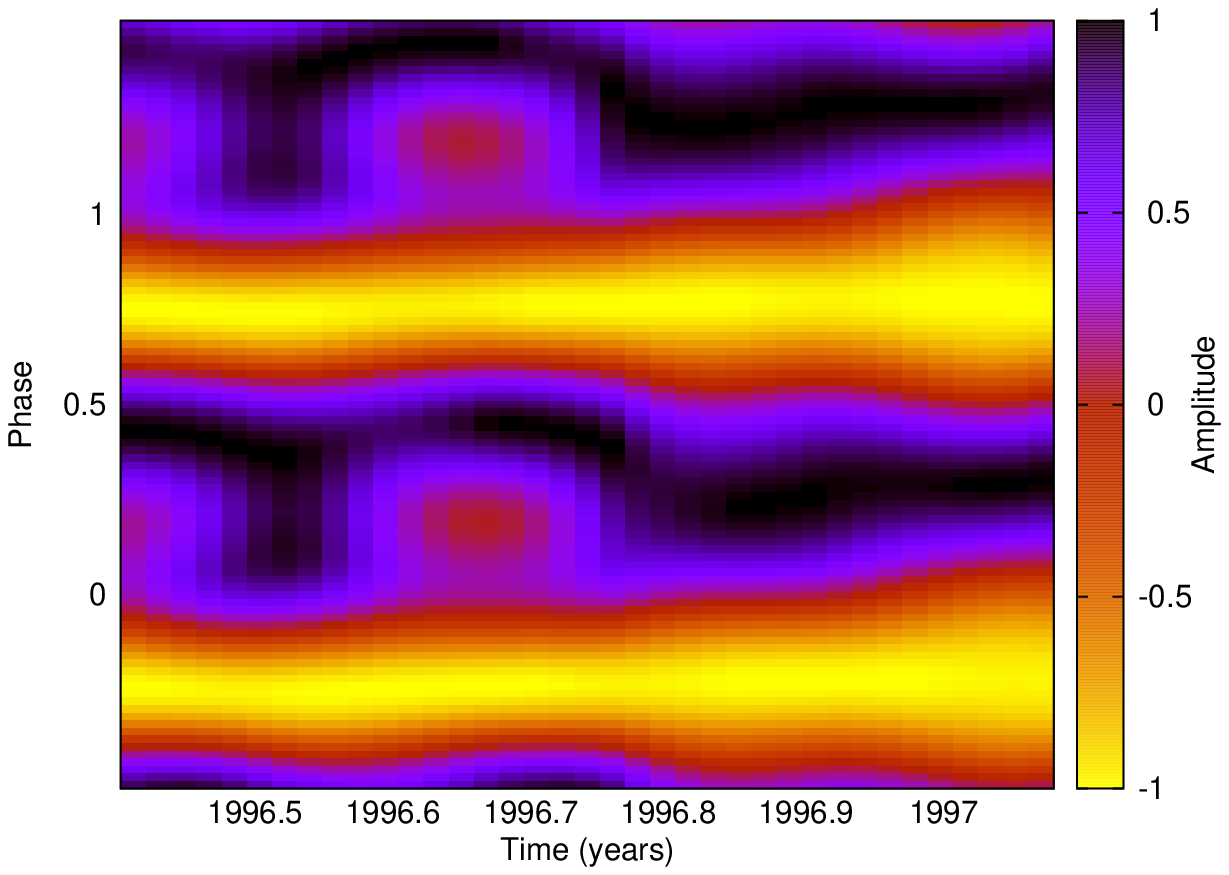}\includegraphics[width=0.33\textwidth]{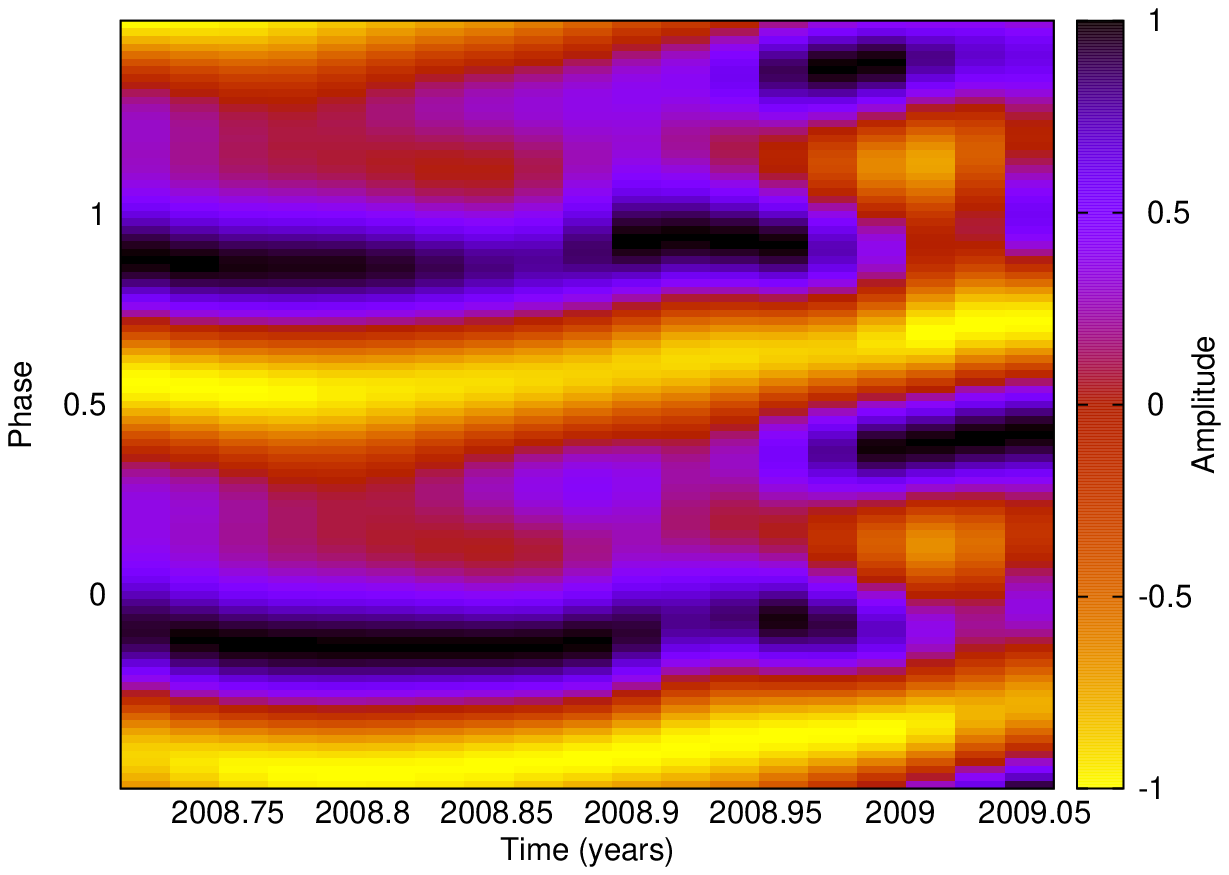}
\caption{Panel in the top row shows SEG12 as an example of a segment with stable phase behavior. Lower rows: Zoom-in to segments in which phase changes were detected. From left to right, top to bottom, SEG07, SEG08, SEG14, SEG15, SEG20 and SEG39.}\label{JUMPS}
\end{center}
\end{figure*}

Next, we divide the data into shorter segments, and perform a 'local'
CF analysis \citep[see also][for similar analysis]{Hackman13}. The
division of the data into segments was done according to the following
rules: We started to group the data from the first point of the
combined data set.  The next point of the data was included in the
same segment, until a gap longer than 60 days was encountered; this
was found to be the absolute maximum for gap length without spoiling
the CF analysis. Then the current segment was considered to be
completed, and a new one started. This way, altogether 42 segments of
varying length and number of data points were formed, but 19 of them
contained too few points to allow for a meaningful CF analysis. The
remaining segments were analyzed with the CF method using $K=2$ and
$L=3$ and utilizing the refined carrier period $P_{\rm spot}$. Most of
the segments show very little variation in phase over time; the
segments can be well described with a horizontal stripe, i.e. the
principal minimum stays at the same phase. As an example of
practically no phase change we show SEG12 in Fig.~\ref{JUMPS}, first
panel; segments resembling this one are indicated with '-' (no change)
in Table~\ref{segments}.

Only four of all the 23 segments containing enough points allowing for
a local CF analysis (SEG07, SEG08, SEG14, SEG39) show clear, abrupt
phase changes. In addition, two of the segments, namely SEG15 and 20,
show disrupted phase behavior, which cannot well be described as phase
jumps (indicated with '?' in Table~\ref{segments}). The two first
segments showing interesting behavior (SEG07 and SEG08) are connected
to the epoch of phase disturbance seen during 1984-1985. Comparing
Figs.~\ref{NEWCARRIER} and \ref{JUMPS} (second and third upper panel)
it seems evident that during 1979-1984, the activity has been
'wobbling' around the 'mean' phase of 0.25, the location of which
roughly agrees with the linear trend, i.e. stays horizontal in
Fig.~\ref{NEWCARRIER}. During 1984-1985, a series of more abrupt phase
jumps are observed to occur, with two short-lived transitions between
phases 0.2 and 0.6. After a third abrupt transition in the middle of
1985, the major part of the activity has remained at the new phase of
0.6 (although hints of a secondary minimum being active at 0.25 can be
seen in the global phase diagram) until 1990, the 'mean' phase of
activity again being consistent with the overall linear trend. Another
phase jump can be seen in SEG14, in the end of the year 1990, when the
main activity seems to have quickly reverted back to the phase of
roughly 0.2, but this spot configuration appears unstable, reverting
back to phase 0.6 via a continuous phase drift during 1992-1995. The
last abrupt phase jump is seen in SEG39, close to the end of the
dataset. Again, a change of roughly 0.5 in phase can be identified.

To conclude, the characteristics of the phase jumps rather closely
resemble what is called the flip-flop \citep{Jetsu93}, i.e. the
activity shifts roughly 0.5 in phase within a time scale of a few
months. The last jump (in SEG39) occurred very near the endpoint of
the time series, due to which it is difficult to conclude about the
persistence of the spot configuration after the event; only one of the
other new spot configurations after a flip-flop type event, seen
during 1985, seems to have persisted over a longer period of time
(several years), while the others have either flipped or drifted back
to the original configurations soon afterwards.

\begin{figure}
\begin{center}
\includegraphics[width=\columnwidth]{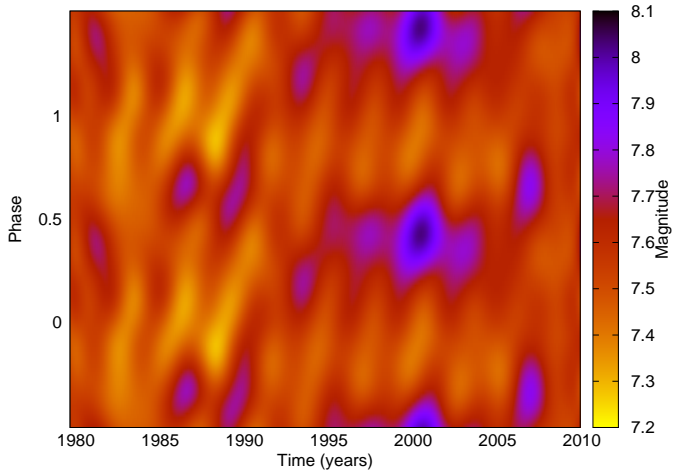}
\caption{CF phase diagram with no
  normalization.}\label{NONNORM}
\end{center}
\end{figure}

There is a ten-year long time interval when the abrupt phase jump
activity clearly ceases to occur (1995-2005), during which only the
linear drift, and some 'wobbling' around the mean phase, is seen. From
spectroscopy \citep[covering the time
1994-2010][]{lindborg11,Hackman12} and spectropolarimetry
\citep[covering the time 2004-2010]{Oleg12} there is some evidence of
the star's magnetic field strength getting weaker and the spots more
randomly distributed over phase during the latest epoch of disturbed
phase activity (2004-2010). The chosen CF visualization scheme of
normalizing each stripe with its own extrema acts to hide possible
amplitude variations in our photometric analysis. Therefore, we
re-compute the phase diagram wrapped with $P_{\rm spot}$ without using
this normalization, the result being shown in Fig.~\ref{NONNORM}. From
this plot it is evident that the photometric minima are indeed deeper
during the period of the most pronounced linear trend, whereas
irregular phase behavior and flip-flop type events are related to the
epoch of weaker photometric minima.  On the other hand, it is
noteworthy that the mean photometric magnitude of the star stays
approximately constant after the year 1990 (see Fig.~\ref{data}),
while only the strength of the variations around the mean are larger
during the epoch with the most pronounced linear trend. The constant
mean level of the photometric magnitude is suggestive of the overall
magnetic activity level of the star having remained unchanged, while
the activity is distributed less axisymmetrically during the epoch of
large magnitude variations.

\begin{table}\label{segments}
\begin{center}
\begin{tabular}{lccc} \hline
Segment &$N$ &$\Delta t$&Events \\ \hline
SEG01   &27  &102.9     &NA      \\
SEG02   &48  &97.0      &NA      \\
SEG03   &60  &130.7     &NA      \\
SEG04   &17  &76.0      &NA      \\
SEG05   &16  &40.0      &NA      \\
SEG06   &43  &88.8      &NA      \\
SEG07   &78  &201.9     &+       \\
SEG08   &98  &167.8     &+       \\
SEG09   &62  &153.5     &NA      \\
SEG10   &39  &87.3      &NA      \\
SEG11   &3   &5.0       &NA      \\
SEG12   &130 &145.2     &-       \\
SEG13   &166 &248.6     &-       \\
SEG14   &85  &237.6     &+       \\
SEG15   &108 &257.6     &?       \\
SEG16   &149 &219.6     &-       \\
SEG17   &180 &248.6     &-       \\
SEG18   &151 &257.6     &-       \\
SEG19   &329 &255.6     &-       \\
SEG20   &359 &259.6     &?       \\
SEG21   &567 &265.6     &-       \\
SEG22   &80  &32.9      &NA      \\
SEG23   &308 &147.8     &-       \\
SEG24   &318 &267.3     &-       \\
SEG25   &219 &246.6     &-       \\	
SEG26   &24  &23.0      &NA      \\
SEG27   &201 &158.1     &-       \\
SEG28   &322 &254.6     &-       \\
SEG29   &49  &46.9      &NA      \\
SEG30   &253 &148.8     &-       \\
SEG31   &58  &41.9      &NA      \\
SEG32   &229 &149.8     &-       \\
SEG33   &285 &260.6     &-       \\
SEG34   &37  &40.0      &NA      \\
SEG35   &196 &138.8     &-       \\
SEG36   &41  &31.0      &NA      \\
SEG37   &172 &131.9     &-       \\
SEG38   &28  &29.0      &NA      \\
SEG39   &163 &136.8     &+       \\
SEG40   &9   &17.0      &NA      \\
SEG41   &48  &130.6     &NA      \\
SEG42   &6   &15.0      &NA      \\ \hline
\end{tabular}
\end{center}
\end{table}

\begin{figure}
\begin{center}
\includegraphics[width=\columnwidth]{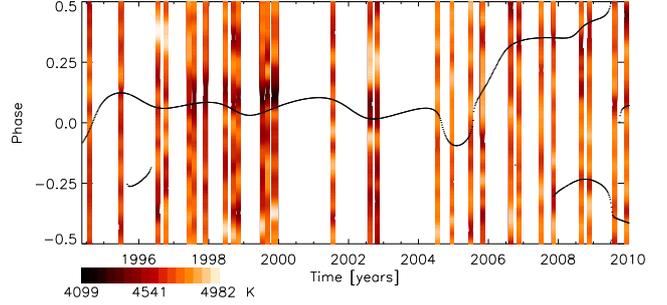}
\caption{Temperature distribution over phase (y-axis) plotted as
  function of time (x-axis) calculated as averages over all latitudes
  from the surface temperature maps obtained with Doppler imaging. The
  black crosses indicate the phases of the photometric minima obtained
  from the CF analysis. The phases are wrapped with the best period
  describing the linear trend, $P_{\rm spot}$ using the first
  timepoint of DI observations as the epoch. }\label{DI_PSPOT}
\end{center}
\end{figure}

\section{Comparison with recent temperature maps}\label{comp}

In this section we compare the temperature minima obtained from
Doppler imaging (DI) by \citet{Hackman11} to the photometric results;
altogether 28 surface temperature maps of the object were derived in
this paper for the years 1994-2010. For comparison purposes, in
Fig.~\ref{DI_PSPOT} we plot the latitudinally averaged temperature
profiles of the DI maps as function of time, but in contrast to
earlier studies \citep[see e.g.][]{Hackman11}, we wrap the phases with
the $P_{\rm spot}$ that was found to describe the linear drift of the
photometric minima. As evident from this figure, one large primary
spot, seen as extensive dark patches in the color contour, was seen on
the object during 1997-2002, the phases of which evolve along rather a
horizontal line when wrapped with $P_{\rm spot}$. During the epochs
1994-1997 and 2004-2010 weaker spot activity, the spots being randomly
distributed over the stellar longitude, is seen. The photometric
minima from our analysis, plotted on top of the DI results with black
crosses, closely coincide with the DI minima for the years with
pronounced spot activity, while the differences are greater for the
other epochs. Comparing to the earlier investigation of
\citet{Hackman11} using a different time series analysis method for
the photometric data, we find the results to be in a fair
agreement. Therefore we conclude that the linear drift pattern is a
robust feature seen both in the DI maps and photometric data analyzed
with different types of methods, persisting at least over a ten year
epoch during 1995-2005. The DI maps, on the other hand, do not show
any clear evidence of the rather continuous phase drift seen during
2005-2009 or the abrupt phase jump detected from photometry in the end
of the dataset (during 2010 in SEG39).

\section{Summary and discussion}\label{concl}

We have collected a 21-year long photometric data set of the
magnetically active primary giant component of the RS CVn binary II
Peg, and analyzed this data set with the CF method, both globally
and locally. As a result, we confirm the earlier results of the spot
activity having been dominated by one primary spotted region almost
through the entire data set, and the existence of a rather persistent
linear drift. Disruptions of the linear trend and complicated phase
behavior are also seen, but the period analysis reveals a 
periodicity with 
$P_{\rm spot}=$6$^{\rm d}$.71086$\pm$0$^{\rm  d}$.00007.
After the linear trend is removed from the data, we
identify several abrupt phase jumps, three of which are analyzed with
the CF method using shorter data segments. These phase jumps closely
resemble what is called the flip-flop event, i.e. the spot activity
changes by roughly 0.5 in phase within months, but only one of the
phase jumps leads to a stable spot configuration persisting over
several years. In other cases, flips back or more continuous drifts
towards the original spot configurations are seen. 

The comparison to Doppler imaging temperature maps confirms the
existence of the linear drift pattern with regular phase behavior for
the epoch 1995-2005, during which the level of spot activity of the
star has been high. During the epochs when photometric analysis shows
disrupted phase behavior and phase jumps, the Doppler images show weak
spots randomly distributed over the longitudes. Therefore, there is
some evidence of the regular drift without phase jumps being related
to the high state, while complex phase behavior and disrupted drift
patterns to the low state of spot activity.  The Zeeman Doppler
imaging results of \cite{Oleg12} indicate the magnetic field strength
of the object getting weaker on average since 2004. The mean
photometric brightness of the object, however, has remained more or
less constant since 1990, suggesting that the overall magnetic
activity level has not changed considerably, although the variations
around the mean have been higher during the epoch of the pronounced
linear trend.

Although the periodicity describing the drift of the primary spot in
the orbital frame of reference of the binary system is stable in
a statistical sense, the trend is clearly disrupted during the epochs of
complex phase behavior; this casts serious doubt of the primary light
curve minimum reflecting the actual, more rapid, rotation of the
primary component of the binary system, that would indicate that the
rotation rates of the stars would still remain de-synchronized by the
mutual tidal forces. Also practically all the dynamo models produced
up to date \citep[see e.g.][]{KR80,MKK99,Moss95,Ilkka02,Mantere13}
show stable drift periods, making it equally hard to explain the
results obtained in this work with the azimuthal dynamo wave
scenario. We note, however, that the recent models of
\citet{Mantere13} indicate that the oscillations in the magnetic
activity level should be connected to the cycle length of the drift of
the non-axisymmetric structures. Using the drift period $P_{\rm spot}$
obtained in this work, an oscillation in the magnetic energy level
with a period of
\begin{equation}
P_{\rm mag}=\frac{P_0 - P_{\rm spot}}{P_0 P_{\rm spot}} \approx 9.2\ {\rm years}
\end{equation}
would be expected. To test this scenario further, however, is out of
the scope of this study.

\newcommand{\etal}{et al.}

\begin{acknowledgements}
  Some of the results presented in this manuscript are based on
  observations made with the Nordic Optical Telescope, operated on the
  island of La Palma jointly by Denmark, Finland, Iceland, Norway, and
  Sweden, in the Spanish Observatorio del Roque de los Muchachos of
  the Instituto de Astrofisica de Canarias. Financial support from the
  Academy of Finland grants No. 112020, 141017 (ML) and 218159 (MJM),
  and financial support from the research programme ``Active Suns'' at
  the University of Helsinki (MJM and TH) are acknowledged.  Astronomy
  at Tennessee State University has been supported by NASA, NSF,
  Tennessee State University, and the State of Tennessee through its
  Centers of Excellence programs.

\end{acknowledgements}

\bibliographystyle{aa}
\bibliography{paper}

\end{document}